\documentstyle[prd,aps]{revtex}

\preprint{LAVAL-PHY-00-13}
\tighten

\begin{document}
\author{L. Marleau and J.-F. Rivard}
\address{D\'epartement de Physique, Universit\'e Laval\\
Qu\'ebec QC Canada, G1K 7P4}
\title{A generating function for all-orders skyrmions}
\date{2000}
\maketitle

\begin{abstract}
We find the generating functions for the Lagrangians of all-orders summable $%
SU(2)$ Skyrmions. We then proceed to construct the explicit form of the
Lagrangian, order by order in the derivatives of the pion field for two
classes of models.
\end{abstract}


\section{Introduction}

According to the $1/N_{c}$ analysis \cite{tHooft74,Witten79}, if an
effective Lagrangian is to represent accurately the low-energy limit of QCD,
it should behave as an effective theory of infinitely many mesons and
include all orders of derivative of the fields. In that sense, the Skyrme
model \cite{Skyrme61} represents a naive (fourth order in derivatives)
attempt to provide such a description. Yet, stable solitonic pion field
configurations (Skyrmions) emerge from this simple model. These solitons are
interpreted as baryons and a number of their properties can be computed
leading to many successful predictions. Unfortunately, the exact form of
such a solution for the low-energy effective Lagrangian is out our reach for
the moment and indeed, would be equivalent to a finding a solution for the
low-energy limit of QCD and perhaps would provide an explanation to
confinement. In the absence of such solutions, one must rely on more
elaborate effective Lagrangians but in doing so one faces problems: (i) the
number of possible extensions, i.e. the number of possible terms at higher
orders becomes increasingly large, (ii) the degree of the equation of motion
becomes arbitrarily large and (iii) adding terms of all orders introduces an
arbitrarily large number of parameters, so one loses any predictive power.
Any one of these problems causes the general approach to become too complex
and its treatment intractable.

A suitable alternative is to impose symmetries or constraints that reduce
this degree of arbitrariness and complexity but still retain the more
interesting features and perhaps introduce new ones. A few years ago, we
proposed such a class of tractable all-orders effective Lagrangians \cite
{Marleau89-10}: Using the hedgehog ansatz for static solution, we required
that the degree of the differential equation for the profile function
remains two. This induces all-orders Skyrmions with a number of interesting
properties: (i) they come from all-orders effective Lagrangians that could
account for infinitely many mesons, (ii) the Lagrangians being chirally
invariant by construction, it allows more control over chiral symmetry
breaking since they can be implemented by hand afterwards, (iii) there
remains some liberty in this class of models so that the physical
constraints can be satisfied properly and predictions can be significantly
improved \cite{Marleau90-13,Jackson91}, (iv) their topological and stability
properties are similar and (v) in some cases new interesting features such
as a two-phase structure \cite{Gustafsson94} arise.

Despite these relative successes, an all-orders Lagrangian written in a
closed form remained to be found. Most of the calculations and constraints
relied on an expression for the energy density of the static hedgehog
solution. A full knowledge of the all-orders Lagrangian is required to
analyze other types of solutions, generalize the model to $SU(3)$ and
hopefully find a link to low-energy QCD. In this work, we construct the most
general all-orders Lagrangians for the class of model introduced in \cite
{Marleau89-10} using a generating function. We then proceed to calculate the
coefficients for the Lagrangian for any arbitrary order. This is done in
Section III after a brief introduction of the all-orders Lagrangian in
Section II. A similar procedure is repeated for effective Lagrangians
induced by hidden gauge symmetry \cite{Bando85} and again, the coefficients
for the Lagrangian are found for any arbitrary order in Section IV. This
construction may have much deeper consequences since it is directly
associated to an underlying gauge theory. Finally, the last section contains
a brief discussion of the results and prospects for further analysis.

\section{More terms to the Skyrme Lagrangian}

Let us recall the Skyrme Lagrangian density for zero pion mass:
\begin{equation}
{\cal L}=-\frac{F_{\pi }^{2}}{16}%
\mathop{\rm Tr}%
L_{\mu }L^{\mu }+\frac{1}{32e^{2}}%
\mathop{\rm Tr}%
f_{\mu \nu }f^{\mu \nu }
\end{equation}
with the notation $L_{\mu }=U^{\dagger }\partial _{\mu }U$ and $f_{\mu \nu
}\equiv [L_{\mu },L_{\nu }]$. The first term, ${\cal L}_{1},$ coincides with
the non-linear $\sigma $-model when one substitutes the degrees of freedom
in the $SU(2)$ matrix $U$ by $\sigma $- and ${\bf \pi }$-fields according to
$U=\frac{2}{F_{\pi }}(\sigma +i{\bf \tau }\cdot {\bf \pi })$. The second
term, ${\cal L}_{2}$, contains higher order derivatives in the pion field
and can account for nucleon-nucleon interactions via pion exchange. ${\cal L}%
_{2}$ was originally added by Skyrme to allow for solitonic solutions. $%
F_{\pi }$ is the pion decay constant (186 MeV) and $e$ is the so-called
Skyrme parameter. Unless $F_{\pi }$ and $e$ are explicitly mentionned, we
shall use more appropriate units in which the Lagrangian rescale as
\begin{equation}
{\cal L}_{1}+\frac{1}{2}{\cal L}_{2}=\left( -\frac{1}{2}%
\mathop{\rm Tr}%
L_{\mu }L^{\mu }\right) +\frac{1}{2}\left( \frac{1}{16}%
\mathop{\rm Tr}%
f_{\mu \nu }f^{\mu \nu }\right)
\end{equation}
and the unit of length is now $\frac{2\sqrt{2}}{eF_{\pi }}$ and the unit of
energy is $\frac{F_{\pi }}{2\sqrt{2}e}.$

In its familiar hedgehog form, the $SU(2)$ matrix $U$ is expressed as
follows:
\[
U({\bf r})=\exp \left[ i{\bf \tau }\cdot \widehat{{\bf r}}F(r)\right]
\]
where $F(r)$ is called the chiral angle or profile function of the solution.
This field configuration constitutes a map from physical space $R^{3}$ onto
the group manifold $SU(2)$ and is assumed to go to the trivial vacuum for
asymptotically large distances. We therefore impose $U(r\rightarrow \infty
)\rightarrow 1$. From this last condition, one may derive the existence of a
topological invariant associated with the mapping. The originality of
Skyrme's idea was to identify this invariant, i.e. the winding number, with
the baryon number.

For a static hedgehog solution, the energy density is given by ${\cal E}=%
{\cal E}_{1}+{\cal E}_{2}$ where the fisrt contribution comes from the
non-linear $\sigma $-model
\begin{equation}
{\cal E}_{1}=-{\cal L}_{1}=-\frac{1}{2}%
\mathop{\rm Tr}%
L_{i}L^{i}=[2a+b]
\end{equation}
with $a\equiv \frac{\sin ^{2}F}{r^{2}}\ \ {\rm and}\ \ b\equiv F^{\prime 2}.$
Although ${\cal L}_{2}$ is quartic in the derivatives of the pion field, $%
{\cal E}_{2}$ adds only a quadratic contribution in $F^{^{\prime }}$ to the
Lagrangian for the static hedgehog solution.
\begin{equation}
{\cal E}_{2}=-{\cal L}_{2}=-\frac{1}{16}%
\mathop{\rm Tr}%
f_{ij}f^{ij}=a[a+2b].
\end{equation}

Despite the relative successes of the Skyrme model, it can only be
considered as a prototype of an effective theory of QCD. For example, there
is no compelling reason or physical grounds for excluding higher order
derivatives in the pion field from the effective Lagrangian. Indeed, large $%
N_{c}$ analysis suggests that bosonization of QCD would most likely involve
an infinite number of mesons. If this is the case, then taking the
appropriate decoupling limits (or large mass limit) for higher spin mesons
leads to an all-orders Lagrangian for pions. One example of higher order
terms is a piece involving $B_{\mu },$ the topological charge density \cite
{Jackson91},
\[
{\cal L}_{J}=c_{J}{\rm Tr}\ [B^{\mu }B_{\mu }]=3a^{2}b\ \ \ {\rm with}\ \ \
B_{\mu }=\epsilon _{\mu \nu \rho \sigma }L^{\nu }L^{\rho }L^{\sigma }
\]
where $c_{J}$ is a constant. It turns out that the term
\begin{equation}
{\cal E}_{3}=-{\cal L}_{3}=\frac{1}{32}%
\mathop{\rm Tr}%
f_{\mu \nu }f^{\nu \lambda }f_{\lambda }^{\ \ \mu }=3a^{2}b
\end{equation}
leads to a similar results.

Several attempts were made to incorporate vector mesons in the Skyrme
picture. Following this approach, we have proposed \cite{Marleau90-13} a
procedure to generalize the Lagrangian density to include all orders in the
derivatives of the pion field in a computationally tractable way. In this
work, we go a step further and will find the exact form of Lagrangian
density at any arbitrary order.

At this point however, it is useful to invoke some relevant links noticed by
Manton \cite{Manton87} between an effective $SU(2)$ scalar Lagrangian and
the strain tensor in the theory of elasticity. As in nonlinear elasticity
theory, the energy density of a Skyrme field depends on the local stretching
associated with the map $U:R^{3}\mapsto S^{3}.$ This is related to the
strain tensor at a point in $R^{3}$ which is defined as
\begin{eqnarray*}
M_{ij} &=&\partial _{i}\Phi \partial _{j}\Phi \ \quad \text{where \quad }%
\Phi =(\sigma ,\pi ^{z},\pi ^{x},\pi ^{y}) \\
&=&-\frac{1}{4}Tr[\{L_{i},L_{j}\}]
\end{eqnarray*}
where $i,j$ refers to the cartesian space coordinates. $M_{ij}$ is a $%
3\times 3$ symmetric matrix with three positive eigenvalues ${\bf X}^{2},%
{\bf Y}^{2},{\bf Z}^{2}$. The vectors ${\bf X,Y,Z}$ are orthogonal here
since they are the principal axes of the strain ellipsoid in this context.
There is a simple geometrical interpretation (due to Manton) for these
objects. They correspond to the changes of length of the images of any
orthogonal system in the space manifold. There are only three fundamental
invariants of $M_{ij}$ with each a simple geometric meaning
\begin{eqnarray*}
Tr[M] &=&{\bf X}^{2}+{\bf Y}^{2}+{\bf Z}^{2}=\sum \left( \text{length}%
\right) ^{2} \\
\frac{1}{2}\left( \left( Tr[M]\right) ^{2}-Tr[M^{2}]\right) &=&{\bf X}^{2}%
{\bf Y}^{2}+{\bf Y}^{2}{\bf Z}^{2}+{\bf Z}^{2}{\bf X}^{2}=\sum \left( \text{%
surface}\right) ^{2} \\
\det M &=&{\bf X}^{2}{\bf Y}^{2}{\bf Z}^{2}=\left( \text{volume}\right) ^{2}
\end{eqnarray*}

Sometimes, it is convenient to use a more general ansatz \cite{Sanyuk93} in
which case the energy contributions to the first three orders take the form

\begin{eqnarray}
{\cal E}_{1} &=&{\bf X}^{2}+{\bf Y}^{2}+{\bf Z}^{2}  \nonumber \\
{\cal E}_{2} &=&\left( {\bf X}\times {\bf Y}\right) ^{2}+\left( {\bf Y\times
Z}\right) ^{2}+\left( {\bf Z}\times {\bf X}\right) ^{2}  \label{XYZ} \\
{\cal E}_{3} &=&3\left( {\bf X}\cdot \left( {\bf Y}\times {\bf Z}\right)
\right) ^{2}  \nonumber
\end{eqnarray}
Since these are the three fundamental invariants in an orthogonal system,
all higher-order Lagrangians can be constructed out of these invariants.
Another important quantity can also be written in term of these objects, the
topological charge density
\[
{\cal Q}=-\frac{1}{2\pi ^{2}}{\bf X}\cdot \left( {\bf Y}\times {\bf Z}%
\right)
\]

For the hedgehog ansatz ${\bf X\cdot Y=Y\cdot Z=X\cdot Z}=0$ and ${\bf X}%
^{2}={\bf Z}^{2}$%
\begin{equation}
a={\bf X}^{2}=\frac{\sin ^{2}F}{r^{2}}\quad b={\bf Y}^{2}=F^{\prime 2}\quad
c={\bf Z}^{2}=\frac{\sin ^{2}F}{r^{2}}  \label{abc}
\end{equation}
and
\begin{eqnarray}
{\cal E}_{1} &=&a+b+c=2a+b  \nonumber \\
{\cal E}_{2} &=&ab+bc+ca=a(a+2b)  \label{e1e2e3} \\
{\cal E}_{3} &=&3abc=3a^{2}b  \nonumber
\end{eqnarray}
The angular integration is trivial in this case.

Let us consider the Lagrangian of an all-orders Skyrme-like model. In
general, it contains even powers of the left-handed current $L_{\mu }$, but
in an orthogonal system, the static energy density is always a combination
of the three invariants e.g. ${\cal E}_{1},{\cal E}_{2}$ and ${\cal E}_{3}.$
It turns out that one can construct a special class \cite{Marleau89-10} of
models whose energy density ${\cal E}$ is at most linear in $b$ (or of
degree two in derivatives of $F$). The static energy density coming from the
Lagrangian of order $2m$ in derivatives of the field is of the form
\begin{equation}
{\cal E}_{m}=a^{m-1}[3a+m(b-a)]  \label{Em}
\end{equation}
for the hedgehog ansatz. The full Lagrangian leads to
\[
{\cal E}=\sum_{m=1}^{\infty }h_{m}{\cal E}_{m}=3\chi (a)+(b-a)\chi ^{\prime
}(a)
\]
where $\chi (x)=\sum_{m=1}^{\infty }h_{m}x^{m}$ and $\chi ^{\prime }(x)=%
\frac{d\chi }{dx}$ and to a profile equation which is computationally
tractable since it is of degree two. In some cases, it is more appropriate
to construct Lagrangians as $m$ powers of the commutators $f_{\mu \nu
}\equiv [L_{\mu },L_{\nu }];$ this leads to vanishing energy density ${\cal E%
}_{2m}$ for $m$ odd $\geq 5$. Jackson et al \cite{Jackson91} found an
elegant expression for the total energy density in terms of the vectors $%
{\bf X,Y,Z}$
\begin{equation}
{\cal E}_{J}=\frac{(a-b)^{3}\chi (c)+(b-c)^{3}\chi (a)+(c-a)^{3}\chi (b)}{%
(a-b)(b-c)(c-a)}  \label{Ejackson}
\end{equation}
where $a,b,c$ are defined in (\ref{abc}). Note that both the numerator and
the denominator are antisymmetric in $a,b,c$ but the total expression is
symmetric as should be expected since ${\cal E}$ is a combination of the
three invariants ${\cal E}_{1},{\cal E}_{2}$ and ${\cal E}_{3}.$ As far as
we know however, there seems to be no fundamental or geometric grounds that
would justify such a form. Moreover, (\ref{Ejackson}) is not very practical
for our purposes since it cannot be easily converted to an expression in
terms of the three invariants ${\cal E}_{1},{\cal E}_{2}$ and ${\cal E}_{3}.$

The mass of the soliton is then written as:

\[
M_{S}=4\pi (\frac{F_{\pi }}{2\sqrt{2}e})\int_{0}^{\infty }r^{2}dr[3\chi
(a)+(b-a)\chi ^{\prime }(a)]
\]
Using the same notation, the chiral equation becomes:

\[
0=\chi ^{\prime }(a)[F^{\prime \prime }+2\frac{F^{\prime }}{r}-2\frac{\sin
F\cos F}{r^{2}}]+a\chi ^{\prime \prime }(a)[-2\frac{F^{\prime }}{r}%
+F^{\prime 2}\frac{\cos F}{\sin F}+\frac{\sin F\cos F}{r^{2}}].
\]
with $a\equiv \frac{\sin ^{2}F}{r^{2}}.$ The Skyrme Lagrangian corresponds
to the case $\chi (a)=\chi _{S}(a)\equiv a+\frac{1}{2}a^{2}$. In the absence
of an exact solution for QCD that would provide a link to a Skyrme-like
Lagrangian, one could consider the toy models in the form of an exponential
or a truncated geometric series:
\[
\chi _{I}(a)=e^{a}-1
\]
\[
\chi _{II,M}(a)=a\frac{1-a^{M}}{1-a}=a+a^{2}+a^{3}+...+a^{M}
\]
which corresponds to the choice $h_{m\leq M}=\frac{(-)^{m-1}}{m}$. Yet,
requiring that a unique soliton solution exists, $\chi (x)$ must satisfy
\begin{eqnarray*}
\frac{d}{dx}\chi (x) &\geq &0,\quad x\geq 0 \\
\frac{d}{dx}\left( \frac{\chi (x)}{x^{3}}\right) &\leq &0,\quad x\geq 0 \\
\frac{d}{dx}\left( \frac{1}{x^{2}}\frac{d}{dx}\chi (x)\right) &\leq &0,\quad
x\geq 0
\end{eqnarray*}
which lead to more physically motivated alternative models due to Jackson et
al \cite{Jackson91} and to Gustaffson and Riska \cite{Gustafsson94}:
\begin{eqnarray}
\chi _{III}(a) &=&\ln (1+a)+\frac{1}{2}a^{2}  \nonumber \\
\chi _{IV}(a) &=&\frac{1}{4}[1-e^{-2a}]+\frac{1}{2}a+\frac{1}{2}a^{2}
\nonumber \\
\chi _{V}(a) &=&a+\frac{a^{3}}{3+2a}  \label{models} \\
\chi _{VI}(a) &=&a+\frac{a^{3}}{3+4a}+\frac{a^{4}}{1+4a^{2}}  \nonumber \\
\chi _{VII,M}(a) &=&a+\frac{a^{M}}{M\sqrt{1+c_{M}a^{2M-6}}}\text{ with }%
c_{M}=\text{ constant.}  \nonumber
\end{eqnarray}
whose phenomenological implications can differ significantly from the
original Skyrme model.

It should be emphasized that although this class of tractable Lagrangians
was originally defined assuming the hedgehog ansatz (i.e. spherically
symmetric solution), the same conditions apply for any orthogonal system
(i.e. ${\bf X\cdot Y=Y\cdot Z=X\cdot Z}=0$) as long as any two of the three
invariants are equal in which case one can write $a={\bf Z}^{2}={\bf X}^{2},$
$b={\bf Y}^{2}$ or $a={\bf Y}^{2}={\bf Z}^{2},$ $b={\bf X}^{2}$ or $a={\bf X}%
^{2}={\bf Y}^{2},$ $b={\bf Z}^{2}$. One can easily construct such a solution
by making a conformal transformation on the hedgehog ansatz for example.

\section{Recursion relation and generating function}

As was pointed out by Jackson et al. in ref. \cite{Jackson91}, ${\cal E}_{m}$
is a function of the three invariants ${\cal E}_{1},{\cal E}_{2}$ and ${\cal %
E}_{3}$ which obeys the recursion relation
\begin{equation}
{\cal E}_{m}={\cal E}_{m-1}{\cal E}_{1}-{\cal E}_{m-2}{\cal E}_{2}+\frac{1}{3%
}{\cal E}_{m-3}{\cal E}_{3}  \label{recenergy}
\end{equation}
Extending this result, it is easy to see that a similar relation holds for
the Lagrangians (since ${\cal L}_{m}\rightarrow -{\cal E}_{m}$ in the static
limit)
\begin{equation}
{\cal L}_{m}=-{\cal L}_{m-1}{\cal L}_{1}+{\cal L}_{m-2}{\cal L}_{2}-\frac{1}{%
3}{\cal L}_{m-3}{\cal L}_{3}  \label{reclag}
\end{equation}
At this point, we must stress on the importance of being able to construct
the Lagrangian ${\cal L}_{m}$ at an arbitrary order $2m$. First the
Lagrangian ${\cal L}_{m}$ is evidently a more fundamental object than the
energy density ${\cal E}_{m}$ since it defines how the scalar fields
interact with each other. Secondly, the recursion relation (\ref{recenergy})
only holds for the hedgehog ansatz but it is sufficient to impose the
condition (\ref{reclag}) on the Lagrangians. Finally, once the Lagrangian is
known at any arbitrary order, it is possible to find other solitonic
solutions, compute their time dependence and, generalize the model from $%
SU(2)$ to $SU(3)$ case or examine other extensions of the Skyrme model.

Let us rewrite the recursion relation (\ref{reclag}) as
\begin{equation}
u_{m}=-u_{m-1}u_{1}+u_{m-2}u_{2}-\frac{1}{3}u_{m-3}u_{3}  \label{recu}
\end{equation}
with $u_{m}={\cal L}_{m}$ for any integer $m.$

By iteration, one would obtain
\begin{eqnarray}
u_{m} &=&-(-u_{m-2}u_{1}+u_{m-3}u_{2}-\frac{1}{3}%
u_{m-4}u_{3})u_{1}+u_{m-2}u_{2}-\frac{1}{3}u_{m-3}u_{3}  \nonumber \\
&=&u_{m-2}(u_{1}^{2}+u_{2})+u_{m-3}(-u_{1}u_{2}-\frac{1}{3}u_{3})+u_{m-4}(%
\frac{1}{3}u_{1}u_{3}) \\
&=&u_{m-3}(-u_{1}^{3}-2u_{1}u_{2}-\frac{1}{3}%
u_{3})+u_{m-4}(u_{1}^{2}u_{2}+u_{2}^{2}+\frac{1}{3}u_{1}u_{3})+u_{m-5}(-%
\frac{1}{3}u_{1}^{2}u_{3}-\frac{1}{3}u_{3}u_{2})  \nonumber \\
&=&...  \nonumber
\end{eqnarray}
and so on. There is several convenient ways to reformulate this recursion
relation e.g. $u_{m}$ can be rewritten as the following matrix operation:
\[
u_{m}=%
\mathop{\rm Tr}%
[T^{m}S_{0}]
\]
with
\begin{equation}
T=\left(
\begin{array}{lll}
-u_{1} & 1 & 0 \\
u_{2} & 0 & 1 \\
-\frac{1}{3}u_{3} & 0 & 0
\end{array}
\right) \ \quad \text{and}\quad S_{0}=3\left(
\begin{array}{ccc}
-1 & -\frac{u_{2}}{u_{3}} & -\frac{u_{1}}{u_{3}} \\
-u_{1} & -\frac{u_{1}u_{2}}{u_{3}} & -\frac{u_{1}^{2}}{u_{3}} \\
u_{2} & \frac{u_{2}^{2}}{u_{3}} & \frac{u_{1}u_{2}}{u_{3}}
\end{array}
\right)
\end{equation}
and the Lagrangian reads
\[
{\cal L=}\sum_{m=1}^{\infty }h_{m}{\cal L}_{m}=%
\mathop{\rm Tr}%
[\chi (T)S_{0}]
\]
But this iterative process is not practical for an arbitrary order $m.$

We need to find a closed form for
\[
u_{n}=\sum_{n_{1},n_{2},n_{3}=0}^{\infty
}C_{n_{1},n_{2},n_{3}}u_{1}^{n_{1}}u_{2}^{n_{2}}u_{3}^{n_{3}}
\]
with $n=n_{1}+2n_{2}+3n_{3}.$ For that purpose, we introduce the generating
function
\begin{equation}
G(u_{1},u_{2},u_{3};x)\equiv c_{1}+\sum_{m=1}^{\infty }u_{m}x^{_{m}}
\label{Gen}
\end{equation}
where accordingly $u_{m}=\left. \frac{1}{m!}\frac{d^{m}}{dx^{m}}%
G(u_{1},u_{2},u_{3};x)\right| _{x=0}$. Here $x$ is an auxiliary variable
introduced for calculational purposes but one could interpret this variable
as a scaling factor in the Skyrme model. Indeed, under a scale
transformation $r\rightarrow \lambda r$, the Lagrangian $u_{m}$ which is of
order $2m$ in derivatives scales as
\[
u_{m}\rightarrow \frac{u_{m}}{\lambda ^{2m}}=u_{m}x^{m}
\]
for $x=\lambda ^{-2}$.

Using the recursion formula, it is easy to find an expression for $G:$ In (%
\ref{Gen}), we have
\[
G=\sum_{m=4}^{\infty }\left( -u_{m-1}u_{1}+u_{m-2}u_{2}-\frac{1}{3}%
u_{m-3}u_{3}\right) x^{m}+u_{3}x^{3}+u_{2}x^{2}+u_{1}x+c_{1}
\]
which upon substituting the summations by the generating function leads to
\[
G=\frac{u_{1}^{2}x^{2}+u_{1}xc_{1}-u_{2}x^{2}c_{1}+\frac{1}{3}%
x^{3}u_{3}c_{1}+u_{3}x^{3}+u_{2}x^{2}+u_{1}x+c_{1}}{\left(
1+u_{1}x-u_{2}x^{2}+\frac{1}{3}u_{3}x^{3}\right) }
\]
This last identity is valid as long as the term in the numerator can be
expanded in powers of $u_{1}x,$ $u_{2}x^{2}$ and $u_{3}x^{3}$. Using the
multinomial expansion
\[
\left( 1+u_{1}x-u_{2}x^{2}+\frac{1}{3}u_{3}x^{3}\right)
^{-1}=\sum_{n_{1},n_{2},n_{3}=0}^{\infty }\frac{(n_{1}+n_{2}+n_{3})!}{%
n_{1}!\ n_{2}!\ n_{3}!}\frac{\left( -\right) ^{n_{1}+n_{3}}}{3^{n_{3}}}%
u_{1}^{n_{1}}u_{2}^{n_{2}}u_{3}^{n_{3}}x^{n_{1}+2n_{2}+3n_{3}}
\]
one gets

\[
G(u_{1},u_{2},u_{3};x)=\sum_{n_{1},n_{2},n_{3}=0}^{\infty }\frac{%
(n_{1}+n_{2}+n_{3})!}{n_{1}!\ n_{2}!\ n_{3}!}\frac{\left( -\right)
^{n_{1}+n_{3}}}{3^{n_{3}}}\frac{\left(
-4n_{1}n_{3}+n_{2}^{2}-n_{2}-2n_{2}n_{3}-3n_{3}^{2}+3n_{3}\right) }{%
(n_{1}+n_{2}+n_{3})(n_{1}+n_{2}+n_{3}-1)}%
u_{1}^{n_{1}}u_{2}^{n_{2}}u_{3}^{n_{3}}x^{n_{1}+2n_{2}+3n_{3}}
\]
This result is independent of the choice of $c_{1}.$

The next step consists in finding the explicit expression for $u_{m}.$ This
is acheived by isolating the term of order $x^{m}$ in the previous
expression. One finds for $m=n_{1}+2n_{2}+3n_{3}\geq 4$%
\[
u_{m}=\sum_{n_{2}=0}^{[\frac{m}{2}]}\sum_{n_{3}=0}^{[\frac{m-2n_{2}}{3}%
]}C_{m-2n_{2}-3n_{3},n_{2},n_{3}}\ u_{1}^{m-2n_{2}-3n_{3}}\ u_{2}^{n_{2}}\
u_{3}^{n_{3}}
\]
where $[z]$ stands for the integer part of $z$ and

\[
C_{n_{1},n_{2},n_{3}}=\frac{(n_{1}+n_{2}+n_{3}-2)!}{n_{1}!\ n_{2}!\ n_{3}!}%
\frac{\left( -\right) ^{n_{1}+n_{3}}}{3^{n_{3}}}\left(
-4n_{1}n_{3}+n_{2}^{2}-n_{2}-2n_{2}n_{3}-3n_{3}^{2}+3n_{3}\right)
\]
The coefficients $C_{n_{1}n_{2}n_{3}}$ for $m=n_{1}+2n_{2}+3n_{3}<4$ are
easy to find by inspection
\begin{eqnarray*}
C_{1,0,0} &=&C_{0,1,0}=C_{0,0,1}=1 \\
C_{0,0,0} &=&C_{2,0,0}=C_{3,0,0}=0
\end{eqnarray*}

Summing up, the full all-orders Lagrangian giving an energy density at most
quadratic in $F^{\prime }$ has the form
\begin{equation}
{\cal L=}\sum_{m=1}^{\infty }h_{m}{\cal L}_{m}=\sum_{m=1}^{\infty
}\sum_{n_{2}=0}^{[\frac{m}{2}]}\sum_{n_{3}=0}^{[\frac{m-2n_{2}}{3}%
]}h_{m}C_{m-2n_{2}-3n_{3},n_{2},n_{3}}\ {\cal L}_{1}^{m-2n_{2}-3n_{3}}\
{\cal L}_{2}^{n_{2}}\ {\cal L}_{3}^{n_{3}}.  \label{fulllag1}
\end{equation}

\section{Hidden gauge symmetry}

In this section, we examine the construction of a generalized Skyrme
Lagrangian based on the hidden gauge symmetry (HGS) formalism \cite
{Bando85,Marleau89-10,Marleau93-11}. The procedure consist in introducing
higher-order gauge terms to the Lagrangian to describe free vector mesons
(for example $SU(2)$ gauge field kinetic Lagrangian $-\frac{1}{4}%
\mathop{\rm Tr}%
F_{\mu \nu }F^{\mu \nu }$) and in the substitution of the derivative by a
covariant derivative to account for scalar-vector interactions. For $SU(2)$
chiral symmetry, the HGS formalism is based on the $SU(2)_{L}\otimes
SU(2)_{R}$ $\otimes $ $SU(2)_{V}$ manifold where $SU(2)_{V}$ is gauged. The
most general Lagrangian involving only two field derivatives is expressed as
\begin{equation}
{\cal L}_{1}^{HGS}=-\frac{F_{\pi }^{2}}{16}[Tr\left( L^{\dagger }D_{\mu
}L-R^{\dagger }D_{\mu }R\right) ^{2}-\alpha Tr\left( L^{\dagger }D_{\mu
}L+R^{\dagger }D_{\mu }R\right) ^{2}]
\end{equation}
where $L(x)\in SU(2)_{L}$ and $R(x)\in SU(2)_{R\text{ }}$. $D_{\mu }$ stands
for the covariant derivative $\partial _{\mu }-igV_{\mu }^{k}\cdot \frac{%
\tau ^{k}}{2}$ where $V_{\mu }$ is the hidden gauge field. When the gauge
vector field is dynamical then a second piece ${\cal L}_{2}^{V}=-\frac{1}{4}%
\mathop{\rm Tr}%
F_{\mu \nu }F^{\mu \nu }$ is added to the Lagrangian. The vector boson $V$
acquires its mass from the same mechanism as the standard gauge bosons with
result $m_{V}^{2}=4\alpha g^{2}F_{\pi }^{2}$. In the large mass limit of the
vector mesons, they decouple and an effective self-interaction for scalar
mesons arise as $F_{\mu \nu }=[D_{\mu },D_{\nu }]\rightarrow f_{\mu \nu
}\equiv [L_{\mu },L_{\nu }]$. Finally, in that limit, ${\cal L}_{1}^{HGS} $
becomes the non-linear $\sigma $-model ${\cal L}_{1}$ whereas ${\cal L}%
_{2}^{V}$ coincides with the Skyrme term ${\cal L}_{2}$.

Following this approach, we have proposed to represent contributions of
order $2n$ in the derivatives of the pion field in terms of the trace of a
product of $n$ $f_{\mu \nu }$'s. Such terms would presumably come from gauge
invariant quantities involving a similar expression with $n$ field strengths
$F_{\mu \nu }$ and would describe exchanges of higher spin mesons. For
example, the lowest-order gauge invariant contributions may have the form
\begin{eqnarray*}
\mathop{\rm Tr}%
F_{\mu \nu }F^{\mu \nu } &\rightarrow &%
\mathop{\rm Tr}%
f_{\mu \nu }f^{\mu \nu } \\
\mathop{\rm Tr}%
F_{\mu }^{\ \nu }F_{\nu }^{\ \lambda }F_{\lambda }^{\ \mu } &\rightarrow &%
\mathop{\rm Tr}%
f_{\mu }^{\ \nu }f_{\nu }^{\ \lambda }f_{\lambda }^{\ \mu } \\
\mathop{\rm Tr}%
\left( F_{\mu \nu }F^{\mu \nu }\right) ^{2} &\rightarrow &%
\mathop{\rm Tr}%
\left( f_{\mu \nu }f^{\mu \nu }\right) ^{2} \\
\mathop{\rm Tr}%
F_{\mu }^{\ \nu }F_{\nu }^{\ \lambda }F_{\lambda }^{\ \sigma }F_{\sigma }^{\
\mu } &\rightarrow &%
\mathop{\rm Tr}%
f_{\mu }^{\ \nu }f_{\nu }^{\ \lambda }f_{\lambda }^{\ \sigma }f_{\sigma }^{\
\mu },\quad \quad {\it etc...}
\end{eqnarray*}
The choice of such combinations is motivated by the possibility that they
could be induced by hidden gauge symmetry (HGS) terms but they also
correspond to exchanges of higher spin particles. They are automatically
chirally invariant. Chiral symmetry breaking must be introduced
independently, usually by adding a pion mass term, which means that in
principle, one has more control on the symmetry breaking mechanism. As we
can see from the first of the above expressions, the Skyrme term itself
emerges from the gauge field kinetic term in the limit of large gauge vector
mass in this formulation.

Furthermore, one can construct a special class \cite{Marleau89-10} of such
combinations which is at most linear in $b$ (or of degree two in derivatives
of $F$). These Lagrangians are a subset of those described in the previous
section. They give a very simple form for the hedgehog energy density
(similar to eq. (\ref{Em}))
\begin{equation}
\widetilde{{\cal E}}_{2m}=a^{2m-1}[3a+2m(b-a)]  \label{E2m}
\end{equation}
where $m$ is an integer. It leads to a chiral angle equation which is
tractable since it is of degree two.

It turns out that for this class of Lagrangians $\widetilde{{\cal E}}_{m}=0$
for $m$ odd $\geq 5$ so we only need to consider Lagrangians which are of
order $4m$ in derivatives of the pion field and have the form $\widetilde{%
{\cal L}}_{2m}$ $\sim
\mathop{\rm Tr}%
(f_{\mu \nu })^{2m}$. Although the constraints look similar, the Lagrangian $%
\widetilde{{\cal L}}_{2m}$ is different from a Lagrangian of the same order
in derivatives ${\cal L}_{2m}$ described in the previous section since the
latter being a combination of ${\cal L}_{1}\sim Tr\left( L_{i}L_{i}\right) $
whereas $\widetilde{{\cal L}}_{2m}$ only involves $f_{\mu \nu }$'s$.$ Yet it
possible to write a recursion relation similar to (\ref{recenergy}) for the
static energies
\[
\widetilde{{\cal E}}_{2m}=\widetilde{{\cal E}}_{2m-2}\widetilde{{\cal E}}%
_{2}-\widetilde{{\cal E}}_{2m-4}\widetilde{{\cal E}}_{4}+\frac{1}{3}%
\widetilde{{\cal E}}_{2m-6}\widetilde{{\cal E}}_{6}
\]
in terms of the three fundamental invariants $\widetilde{{\cal E}}_{2},%
\widetilde{{\cal E}}_{4}$ and $\widetilde{{\cal E}}_{6}.$ $\widetilde{{\cal E%
}}_{2m}$ arises from the Lagrangian $\widetilde{{\cal L}}_{2m}$ which obeys
\[
\widetilde{{\cal L}}_{2m}=-\widetilde{{\cal L}}_{2m-2}\widetilde{{\cal L}}%
_{2}+\widetilde{{\cal L}}_{2m-4}\widetilde{{\cal L}}_{4}-\frac{1}{3}%
\widetilde{{\cal L}}_{2m-6}\widetilde{{\cal L}}_{6}
\]
with
\begin{eqnarray*}
\widetilde{{\cal L}}_{2} &=&\frac{1}{16}%
\mathop{\rm Tr}%
f_{\mu \nu }f^{\mu \nu } \\
\widetilde{{\cal L}}_{4} &=&\frac{1}{64}\left(
\mathop{\rm Tr}%
f_{\mu }^{\ \nu }f_{\nu }^{\ \lambda }f_{\lambda }^{\ \sigma }f_{\sigma }^{\
\mu }-%
\mathop{\rm Tr}%
\{f_{\mu }^{\ \nu },f_{\lambda }^{\ \sigma }\}f_{\nu }^{\ \lambda }f_{\sigma
}^{\ \mu }\right) \\
\widetilde{{\cal L}}_{6} &=&-\frac{1}{256}\left(
\mathop{\rm Tr}%
f_{\mu }^{\ \nu }f_{\nu }^{\ \lambda }f_{\lambda }^{\ \sigma }f_{\sigma }^{\
\rho }f_{\rho }^{\ \omega }f_{\omega }^{\ \mu }-2%
\mathop{\rm Tr}%
\{f_{\mu }^{\ \nu },f_{\lambda }^{\ \sigma }\}f_{\nu }^{\ \lambda }f_{\sigma
}^{\ \rho }f_{\rho }^{\ \omega }f_{\omega }^{\ \mu }+\frac{3}{2}%
\mathop{\rm Tr}%
\{f_{\mu }^{\ \nu },f_{\lambda }^{\ \sigma }\}\{f_{\nu }^{\ \lambda
},f_{\rho }^{\ \omega }\}f_{\sigma }^{\ \rho }f_{\omega }^{\ \mu }\right) .
\end{eqnarray*}

Again we are interested in a closed form for $\widetilde{{\cal L}}_{2m}$ at
any arbitrary order, i.e.
\[
\widetilde{{\cal L}}_{2m}=\sum_{n_{1},n_{2},n_{3}=0}^{\infty
}C_{n_{1},n_{2},n_{3}}\ \widetilde{{\cal L}}_{2}^{n_{1}}\ \widetilde{{\cal L}%
}_{4}^{n_{2}}\ \widetilde{{\cal L}}_{6}^{n_{3}}
\]
with $m=n_{1}+2n_{2}+3n_{3}.$ Using the generating function technique, we
see that the procedure is identical to that in the Section III upon the
substitution $u_{m}=\widetilde{{\cal L}}_{2m}$ for any integer $m.$ The full
all-orders Lagrangian in derivatives of the pion field give rise to an
energy density at most quadratic in $F^{\prime }$ and has the form
\begin{eqnarray}
\widetilde{{\cal L}} &=&{\cal L}_{1}+\sum_{m=1}^{\infty }h_{2m}\widetilde{%
{\cal L}}_{2m}  \label{fulllag2} \\
&=&{\cal L}_{1}+\sum_{m=1}^{\infty }\sum_{n_{2}=0}^{[\frac{m}{2}%
]}\sum_{n_{3}=0}^{[\frac{m-2n_{2}}{3}]}h_{2m}C_{m-2n_{2}-3n_{3},n_{2},n_{3}}%
\ \widetilde{{\cal L}}_{2}^{m-2n_{2}-3n_{3}}\ \widetilde{{\cal L}}%
_{4}^{n_{2}}\ \widetilde{{\cal L}}_{6}^{n_{3}}.  \nonumber
\end{eqnarray}
where the first term is the nonlinear sigma model and the remaining of the
expression accounts for the Skyrme term and higher-order terms. In this
class of models, contributions of order $4m+2$ in derivatives of the pion
field are absent.

Following the HGS formalism, this Lagrangian corresponds to the large mass
limit of the vector mesons of a class of scalar gauged field theory
described by the Lagrangian
\[
{\cal L}={\cal L}_{1}^{HGS}+\sum_{m=1}^{\infty }h_{2m}{\cal L}_{2m}^{V}
\]
where ${\cal L}_{2m}^{V}$ is obtained from $\widetilde{{\cal L}}_{2m}$ upon
substitution $f_{\mu \nu }\rightarrow F_{\mu \nu }.$ The study of such
theories and their justification based on physical grounds remains to be
addressed.

\section{Conclusion}

Under their respective constraints, both (\ref{fulllag1}) and (\ref{fulllag2}%
) describe the most general all-orders Lagrangians. However some specific
models are worth mentionning. For example, a simple choice of coefficients $%
h_{m}=1$ corresponds to static energy densities
\begin{eqnarray*}
\chi (a) &=&\frac{a}{1-a}=a+a^{2}+a^{3}+... \\
\widetilde{\chi }(a) &=&a+\frac{a^{2}}{1-a^{2}}=a+a^{2}+a^{4}+...
\end{eqnarray*}
respectively and are induced by the Lagrangians written only in terms of the
generating function
\begin{eqnarray}
{\cal L} &=&G({\cal L}_{1},{\cal L}_{2},{\cal L}_{3};1)=\frac{{\cal L}_{1}+(%
{\cal L}_{1}^{2}+{\cal L}_{2})+{\cal L}_{3}}{\left( 1+{\cal L}_{1}-{\cal L}%
_{2}+\frac{1}{3}{\cal L}_{3}\right) }  \label{LCA} \\
\widetilde{{\cal L}} &=&{\cal L}_{1}+G({\cal L}_{2},{\cal L}_{4},{\cal L}%
_{6};1)={\cal L}_{1}+\frac{\widetilde{{\cal L}}_{2}+(\widetilde{{\cal L}}%
_{2}^{2}+\widetilde{{\cal L}}_{4})+\widetilde{{\cal L}}_{6}}{\left( 1+%
\widetilde{{\cal L}}_{2}-\widetilde{{\cal L}}_{4}+\frac{1}{3}\widetilde{%
{\cal L}}_{6}\right) }  \label{LTCA}
\end{eqnarray}
In general, the term in the numerator of the generating function is not
analytic so the expression must be understood as the correct analytic
continuation of its series representation. In both cases, we chose $c_{1}=0$
since it only adds a constant piece to the full Lagrangian and otherwise
would lead to an infinite energy solution. This result is easily generalized
to the model with coefficients $h_{m}=h^{m}$ since it is equivalent to a
scale transformation and may be performed through the change of variables $%
a\rightarrow ha,$ ${\cal L}_{1}\rightarrow {\cal L}_{1}h,$ ${\cal L}%
_{2}\rightarrow {\cal L}_{2}h^{2},$ ${\cal L}_{3}\rightarrow {\cal L}%
_{3}h^{3}$ in the above expressions.

Let us now examine the Lagrangian in expression (\ref{LCA}). For the
hedgehog ansatz, we find an energy density
\begin{equation}
{\cal E}=-\frac{-{\cal E}_{1}+({\cal E}_{1}^{2}-{\cal E}_{2})-{\cal E}_{3}}{%
\left( 1-{\cal E}_{1}+{\cal E}_{2}-\frac{1}{3}{\cal E}_{3}\right) }
\label{Emodel}
\end{equation}
where ${\cal E}_{1},{\cal E}_{2}$ and ${\cal E}_{3}$ are given by (\ref
{e1e2e3}). This energy density ${\cal E}$ is obviously symmetric in $a,b,c$
(for the hedgehog ansatz $a=c)$ yet, it coincides with the energy density $%
{\cal E}_{J}$ given by the formula (\ref{Ejackson}) of Jackson et al, i.e.
the ratio of two antisymmetric expressions. For this rather simple model, it
turns out to be possible to proceed backwards and deduce the full Lagrangian
from ${\cal E}_{J}:$ Starting from
\[
{\cal E}_{J}=\frac{(a-b)^{3}\left( \frac{c}{1-c}\right) +(b-c)^{3}\left(
\frac{a}{1-a}\right) +(c-a)^{3}\left( \frac{b}{1-b}\right) }{(a-b)(b-c)(c-a)}
\]
and collecting terms which scale identically in the numerator and in the
denominator, one can recast the energy density as
\[
{\cal E}_{J}=-\frac{-\left( a+b+c\right) +(a^{2}+ab+ca+b^{2}+bc+c^{2})-3abc}{%
\left( 1-\left( a+b+c\right) +\left( ab+bc+ca\right) -abc\right) }
\]
This is equivalent to (\ref{Emodel}) and suggest that the full Lagrangian
has precisely the form in (\ref{LCA}). A similar procedure can also be used
for models where $\chi (a)$ is a rational function (quotient of two
polynomials in $a$) e.g. $\chi _{II,M}(a),\chi _{V}(a)$ and $\chi _{VI}(a)$
in (\ref{models}) but the complexity of calculations rises as the degree of
the polynomials increases. On the other hand, these models are also easy to
obtain in terms of the generating function $G.$ More elaborate models
requires the general expressions (\ref{fulllag1}) and (\ref{fulllag2}).

It is easy to show ${\cal E}={\cal E}_{J}$ holds for any model represented
by a function $\chi (x)=\sum_{m=1}^{\infty }h_{m}x^{m}$ in (\ref{Ejackson})
using the most general form
\[
{\cal E=}\sum_{m=1}^{\infty }\sum_{n_{2}=0}^{[\frac{m}{2}]}\sum_{n_{3}=0}^{[%
\frac{m-2n_{2}}{3}]}h_{m}C_{m-2n_{2}-3n_{3},n_{2},n_{3}}\ (-1)^{m-n_{2}+1}%
{\cal E}_{1}^{m-2n_{2}-3n_{3}}\ {\cal E}_{2}^{n_{2}}\ {\cal E}_{3}^{n_{3}}
\]
and (\ref{e1e2e3}). Clearly, this result was to be expected since both
constructions are based on the same constraints for the energy density. So
besides finding the complete form of the most general all-orders
Lagrangians, we have come up with a way to write the energy density in terms
of polynomials symmetric in $a,b,c$.

In principle, it is possible to get an expression for the full Lagrangians
written in term of the generating function $G$ only, for any model in this
class, e.g. models defined in (\ref{models}). Note that in some cases, it
may be more convenient to work with the {\em exponential generating function,%
} $E(x)=\sum_{m=1}^{\infty }\frac{1}{m!}u_{m}x^{_{m}}$.

In summary, the generating function $G$ was constructed for the class of
models in which the energy density for the hedgehog ansatz is at most
quadratic in $F^{\prime }$ and therefore computationally tractable. It led
to an explicit expression for the coefficients $C_{n_{1},n_{2},n_{3}}$ in (%
\ref{fulllag1}) and (\ref{fulllag2}). But now that the full Lagrangians are
known, a number of questions remain to be addressed: (a) Solutions other
than the $N=1$ static hedgehog can now be analyzed thoroughly. For example,
one can propose a different or a more general ansatz according to (\ref{XYZ}%
). (b) Also, solutions for $N>1$ skyrmions should be examined either by a
direct numerical calculation or a convenient ansatz (such as rational maps
\cite{Houghton98}). (c) It is now possible to write the time dependence for
this class of models for which perhaps one could eventually provide a proper
description.(d) The all-orders Lagrangians also allow to study analytically
the identity map and to construct the mode spectrum in a closed form.
Finally, the knowledge of this class of Lagrangians allows (e) to extend the
models, for example in an extension from the $SU(2)$ to the $SU(3)$ symmetry
group or (f) to use it in other areas, such as in weak skyrmions, baby
skyrmions, etc...

We would like to thank L. B\'{e}gin for useful discussions. This research
was supported by the Natural Science and Engineering Research Council of
Canada and by the Fonds pour la Formation de Chercheurs et l'Aide \`{a} la
Recherche du Qu\'{e}bec.


\begin{references}
\bibitem{tHooft74}  G. 't Hooft, Nucl. Phys. {\bf B72,} 461 (1974).

\bibitem{Witten79}  E. Witten, Nucl. Phys. {\bf B160}, 57 (1979).

\bibitem{Skyrme61}  T.H.R. Skyrme, Proc.~R.~Soc.~London.~{\bf A2603,} 127
(1961).

\bibitem{Marleau89-10}  L. Marleau, Phys.~Lett.~{\bf B235,} 141 (1990).

\bibitem{Marleau90-13}  S. Dub\'{e} and L. Marleau, Phys.~Rev.~{\bf D41,}
1606 (1990).

\bibitem{Jackson91}  A.D. Jackson, C. Weiss and A. Wirzba, Nucl.~Phys.~{\bf %
A529,} 741 (1991).

\bibitem{Gustafsson94}  K. Gustafsson and D.O. Riska, Nucl. Phys. {\bf A571}%
, 645 (1994).

\bibitem{Bando85}  M. Bando, T. Kugo, S. Uehara, K. Yamawaki and T.
Yanagida, Phys.~Rev.~Lett.~{\bf 54} 1215 (1985).

\bibitem{Manton87}  M.S. Manton, Commun. Math. Phys. {\bf 111, }469 (1987).

\bibitem{Sanyuk93}  V.G. Makhankov, Y.P. Rybakov and V.I. Sanyuk, The SKyrme
Model: Fundamentals, Methods, Applications. Ed. Springler-Verlag, 1993.

\bibitem{Marleau93-11}  L. Marleau and H. Omari, Phys. Rev. {\bf D48},
(1993).

\bibitem{Houghton98}  C.J. Houghton, N.S. Manton and P.M. Sutcliffe, Nucl.
Phys. {\bf B510}, 507 (1998).
\end{references}
\end{document}